\documentstyle[11pt]{article}
\newcommand{\cl}{\centerline}
\renewcommand{\theequation}{\thesection.\arbic{equation}}
\renewcommand{\theequation}{A.\arbic{equation}}

\renewcommand\d{{\rm d}}
\newcommand\beq{\begin{equation}}
\newcommand\eeq{\end{equation}}
\newcommand\bea{\begin{eqnarray}}
\newcommand\eea{\end{eqnarray}}

\begin{document}

\begin{titlepage}
\setlength{\textwidth}{5.0in}
\setlength{\textheight}{7.5in}
\setlength{\parskip}{0.0in}
\setlength{\baselineskip}{18.2pt}
\hfill
SNUTP98/042
\vfill
\cl{\Large{{\bf Strange Form Factors and Sum Rules}}}\par
\cl{\Large{{\bf of Baryon Decuplet}}}\par
\vskip 0.8cm
\cl{Soon-Tae Hong$^{a}$ and Dong-Pil Min$^{b}$}\par
\vskip 0.4cm
\cl{$^{a}$Deapartment of Physics and Basic Science Research Institute}\par
\vskip 0.2cm
\cl{Sogang University, C.P.O. Box 1142, Seoul 100-611, Korea}\par
\vskip 0.2cm
\cl{$^{b}$Deapartment of Physics and Center for Theoretical Physics}\par
\vskip 0.2cm
\cl{Seoul National University, Seoul 151-742, Korea}\par
\vskip 0.8cm
\cl{\today}
%\cl{July 14, 1998}
\vskip 0.8cm
\vfill
\begin{center}
{\bf ABSTRACT}
\end{center}
\begin{quotation}
Treating perturbatively the flavor symmetry breaking effects resided in the 
wave function as well as in the SU(3) Lagrangian, we calculate the strange 
form factors of decuplet baryons.  The higher dimensional representation 
mixing components in the baryon wave function of the deccuplet baryons are 
found to yield the important contributions to the strange form factors, 
especially of the $\Delta$ and $\Sigma^*$ baryons.  The model independent sum 
rules among the baryon decuplet magnetic moments are also derived.
\end{quotation}
\end{titlepage}

\section{Introduction}

\setcounter{equation}{0} \renewcommand{\theequation}{\arabic{section}.%
\arabic{equation}}

Since Coleman and Glashow\cite{cg} predicted the magnetic moments of the
baryon octet about forty years ago, there has been a lot of progress in both
the theoretical paradigm and experimental verification for the baryon
magnetic moments. Recently the measurements of the baryon decuplet magnetic
moments were reported for $\mu_{\Delta^{++}}$\cite{boss} and $%
\mu_{\Omega^{-}}$\cite{die} to yield a new avenue for understanding the
hadron structure. The magnetic moments of baryon decuplet have been
theoretically investigated in several models, {\it e.g.}, in the quenched
lattice gauge theory\cite{lein}, the quark models\cite{sch,lin1}, the chiral
bag model\cite{decup}, the chiral perturbation theory\cite{but}, the chiral
quark soliton model\cite{kim}, the QCD sum rules\cite{lee1,lee2} and the
chiral quark model\cite{lin2}.

On the other hand the chiral and SU(3) flavor symmetry breaking (FSB)
effects in the chiral bag model (CBM)\cite{cbm} are induced by the different
pseudoscalar meson masses and decay constants outside and the quark masses
inside the bag. Especially the SU(3) FSB originates from the strangeness
degrees of freedom in the baryon structure, which has been significantly
discussed after the EMC experiment on deep inelastic muon scattering\cite
{emc} suggested a lingering question.

Quite recently, the SAMPLE collaboration\cite{sample} reported the
experimental data of the proton strange form factor through parity violating
electron scattering\cite{mck89}. Moreover, McKeown\cite{mck} has shown that
the strange form factor of proton should be positive by using the conjecture
that the up-quark effects are generally dominant in the flavor dependence of
the nucleon properties. This result is contrary to the negative values of
the proton strange form factor which result from most of the model
calculations\cite{jaffe,musolf,koepf,park} except that of Hong and Park\cite
{hp} based on the SU(3) CBM\cite{cbm2} and that of Meissner and co-workers 
\cite{meissner} in the heavy baryon chiral perturbation theory. The CBM
prediction on the positive strange form factor of the proton was also
justified\cite{hpm} by adjusting the inertia parameters in a systematic way.

In this paper we will report the result on strange form factor of the baryon
decuplet in the minimal multi-quark structure where the symmetry breaking
mass effects are treated in the perturbative scheme of the CBM, as an
extension of our former work on the baryon octet.\cite{hp,hpm}

In Section 2, the symmetry breaking in the CBM will be discussed in terms of
the physical operators in the adjoint representation to yield the model
independent sum rules. In Section 3, we will introduce the baryon wave
functions in the multiquark Fock space of the higher representation mixing
scheme to obtain the strange form factors of baryon decuplet.

\section{Magnetic moments of baryon decuplet}

\setcounter{equation}{0} \renewcommand{\theequation}{\arabic{section}.%
\arabic{equation}}

In the CBM with the broken U-spin symmetry the Lagrangian is of the form 
\begin{eqnarray}
& &{\cal L}={\cal L}_{CS}+{\cal L}_{CSB}+{\cal L}_{FSB}  \label{lag} \\
& &{\cal L}_{CS}=\bar{\psi}i\gamma^{\mu}\partial_{\mu}\psi -\frac{1}{2}\bar{%
\psi}U_{5}\psi\Delta_{B}  \nonumber \\
& &\ \ \ \ \ \ \ \ \ \ +(-\frac{1}{4}f_{\pi}^{2}{\rm tr}(l_{\mu}l^{\mu}) +%
\frac{1}{32e^{2}}[l_{\mu},l_{\nu}]^{2}+{\cal L}_{WZW})\bar{\Theta}_{B} 
\nonumber \\
& &{\cal L}_{CSB}=-\bar{\psi}M\psi\Theta_{B}+\frac{1}{4}f_{\pi}^{2}m_{%
\pi}^{2} {\rm tr}(U+U^{\dagger}-2)\bar{\Theta}_{B}  \nonumber \\
& &{\cal L}_{FSB}=\frac{1}{6}f_{\pi}^{2}(\chi^{2}m_{K}^{2}-m_{\pi}^{2}) {\rm %
tr}((1-\sqrt{3}\lambda_{8})(U+U^{\dagger}-2))\bar{\Theta}_{B}  \nonumber \\
& &\ \ \ \ \ \ \ \ \ \ -\frac{1}{12}f_{\pi}^{2}(\chi^{2}-1){\rm tr}((1-\sqrt{%
3} \lambda_{8})(Ul_{\mu}l^{\mu}+l_{\mu}l^{\mu}U^{\dagger}))\bar{\Theta}_{B}
\end{eqnarray}
where the quark field $\psi$ has SU(3) flavor degrees of freedom and the
chiral field $U=e^{i\lambda_{a}\pi_{a}/f_{\pi}}\in$ SU(3) is described by
the pseudoscalar meson fields $\pi_{a}$ (a=1,...8) and Gell-Mann matrices $%
\lambda_{a}$ with $\lambda_{a}\lambda_{b}=\frac{2}{3}\delta_{ab}+(if_{abc}
+d_{abc})\lambda_{c}$ and $\Theta_{B}(=1-\bar{\Theta}_{B})$ is the bag theta
function (one inside the bag and zero outside the bag). Here $%
l_{\mu}=U^{\dagger}\partial_{\mu}U$ and ${\cal L}_{WZW}$ stands for the
topological Wess-Zumino-Witten (WZW) term. In the numerical calculation we
will use the parameters $e=4.75$, $f_{\pi}=93$ MeV and $f_{K}=114$ MeV.

Here the chiral symmetry (CS) is broken by the quark masses $M={\rm diag}
(m_{u},m_{d},m_{s})$ and pion mass $m_{\pi}$ in ${\cal L}_{CSB}$.
Furthermore the SU(3) FSB with $m_{K}/m_{\pi}\neq 1$ and $%
\chi=f_{K}/f_{\pi}\neq 1$ is included in ${\cal L}_{FSB}$.

Even though the mass terms in ${\cal L}_{CSB}$ and ${\cal L}_{FSB}$ break
both the SU$_{L}$(3)$\times$SU$_{R}$(3) and diagonal SU(3) symmetry so that
chiral symmetry cannot be conserved, these terms without derivatives yield
no explicit contribution to the electromagnetic (EM) currents $J^{\mu}$ and
at least in the adjoint representation of the SU(3) group the EM currents
are conserved and of the same form as the chiral limit result $J^{\mu}_{CS}$
to preserve the U-spin symmetry. However the derivative-dependent term in $%
{\cal L}_{FSB}$ gives rise to the U-spin symmetry breaking conserved EM
currents $J^{\mu}_{FSB}$ so that $J^{\mu}=J^{\mu}_{CS}+J^{\mu}_{FSB}$.

Assuming that the hedgehog classical solution in the meson phase $U_{0}=
e^{i\lambda_{i}\hat{r}_{i}\theta (r)}$ (i=1,2,3) is embedded in the SU(2)
isospin subgroup of SU(3) and the Fock space in the quark phase is described
by the $N_{c}$ valence quarks and the vacuum structure composed of quarks
filling the negative energy sea, the CBM generates the zero mode with the
collective variable $A(t)\in$ SU(3) by performing the slow rotation $%
U\rightarrow AU_{0} A^{\dagger}$ and $\psi\rightarrow A\psi$ on SU(3) group
manifold.

Given the spinning CBM ansatz, the EM currents yield the magnetic moment
operators $\hat{\mu}^{i}=\hat{\mu}^{i(3)}+\frac{1}{\sqrt{3}}\hat{\mu}^{i(8)}$
where $\hat{\mu}^{i(a)}=\hat{\mu}^{i(a)}_{CS}+\hat{\mu}^{i(a)}_{FSB}$ with 
\begin{eqnarray}
\hat{\mu}^{i(a)}_{CS}&=&-{\cal N}D_{ai}^{8}-{\cal N}^{%
\prime}d_{ipq}D_{ap}^{8} \hat{T}_{q}^{R}+\frac{N_{c}}{2\sqrt{3}}{\cal M}%
D_{a8}^{8}\hat{J}_{i}  \nonumber \\
\hat{\mu}^{i(a)}_{FSB}&=&-{\cal P}D_{ai}^{8}(1-D_{88}^{8})+{\cal Q} \frac{%
\sqrt{3}}{2}d_{ipq}D_{ap}^{8}D_{8q}^{8}
\end{eqnarray}
where ${\cal M}$, ${\cal N}$, ${\cal N}^{\prime}$, ${\cal P}$ and ${\cal Q}$
are the inertia parameters calculable in the CBM.{\bf \cite{hpm} }

Using the theorem that the tensor product of the Wigner D functions can be
decomposed into sum of the single D functions, the isovector and isoscalar
parts of the operator $\hat{\mu}^{i(a)}_{FSB}$ are then rewritten as 
\begin{eqnarray}
\hat{\mu}^{i(3)}_{FSB}&=&{\cal P}(-\frac{4}{5}D_{3i}^{8}+\frac{3}{10}
D_{3i}^{27})+{\cal Q}(\frac{3}{10}D_{3i}^{8}-\frac{3}{10}D_{3i}^{27}) 
\nonumber \\
\hat{\mu}^{i(8)}_{FSB}&=&{\cal P}(-\frac{6}{5}D_{8i}^{8}+\frac{9}{20}
D_{8i}^{27})+{\cal Q}(-\frac{3}{10}D_{8i}^{8}-\frac{9}{20}D_{8i}^{27}).
\end{eqnarray}
Here one notes that the ${\bf 1}$, ${\bf 10}$ and $\bar{{\bf 10}}$
irreducible representations (IRs) do not occur in the decuplet baryons while 
${\bf 10}$ and $\bar{{\bf 10}}$ IRs appear together in the isovector channel
of the baryon octet to conserve the hermiticity of the operator.

With respect to the decuplet baryon wave function $\Phi_{B}^{\lambda}= \sqrt{%
{\rm dim}(\lambda)}D_{ab}^{\lambda}$ with the quantum numbers $a=(Y;I,I_{3})$
($Y$; hypercharge, $I$; isospin) and $b=(Y_{R};J,-J_{3})$ ($Y_{R}$; right
hypercharge, $J$; spin) and $\lambda$ the dimension of the representation,
the spectrum of the magnetic moment operator $\hat{\mu}^{i}$ has the
following hyperfine structure in the adjoint representation 
\begin{eqnarray}
\mu_{\Delta^{++}}&=&\frac{1}{8}{\cal M}+\frac{1}{2}({\cal N}-\frac{1} {2%
\sqrt{3}}{\cal N}^{\prime})+\frac{3}{7}{\cal P}-\frac{3}{56}{\cal Q} 
\nonumber \\
\mu_{\Delta^{+}}&=&\frac{1}{16}{\cal M}+\frac{1}{4}({\cal N}-\frac{1} {2%
\sqrt{3}}{\cal N}^{\prime})+\frac{5}{21}{\cal P}+\frac{1}{84}{\cal Q} 
\nonumber \\
\mu_{\Delta^{0}}&=&\frac{1}{21}{\cal P}+\frac{13}{168}{\cal Q}  \nonumber \\
\mu_{\Delta^{-}}&=&-\frac{1}{16}{\cal M}-\frac{1}{4}({\cal N}-\frac{1} {2%
\sqrt{3}}{\cal N}^{\prime})-\frac{1}{7}{\cal P}+\frac{1}{7}{\cal Q} 
\nonumber \\
\mu_{\Sigma^{*+}}&=&\frac{1}{16}{\cal M}+\frac{1}{4}({\cal N}-\frac{1} {2%
\sqrt{3}}{\cal N}^{\prime})+\frac{19}{84}{\cal P}-\frac{17}{168}{\cal Q} 
\nonumber \\
\mu_{\Sigma^{*0}}&=&\frac{1}{84}{\cal P}-\frac{1}{84}{\cal Q}  \nonumber \\
\mu_{\Sigma^{*-}}&=&-\frac{1}{16}{\cal M}-\frac{1}{4}({\cal N}-\frac{1} {2%
\sqrt{3}}{\cal N}^{\prime})-\frac{17}{84}{\cal P}+\frac{13}{168}{\cal Q} 
\nonumber \\
\mu_{\Xi^{*0}}&=&-\frac{1}{42}{\cal P}-\frac{17}{168}{\cal Q}  \nonumber \\
\mu_{\Xi^{*-}}&=&-\frac{1}{16}{\cal M}-\frac{1}{4}({\cal N}-\frac{1} {2\sqrt{%
3}}{\cal N}^{\prime})-\frac{11}{42}{\cal P}+\frac{1}{84}{\cal Q}  \nonumber
\\
\mu_{\Omega^{-}}&=&-\frac{1}{16}{\cal M}-\frac{1}{4}({\cal N}-\frac{1} {2%
\sqrt{3}}{\cal N}^{\prime})-\frac{9}{28}{\cal P}-\frac{3}{56}{\cal Q}.
\label{dec}
\end{eqnarray}

In the SU(3) flavor symmetric limit with the chiral symmetry breaking masses 
$m_{u}=m_{d}=m_{s}$, $m_{K}=m_{\pi}$ and decay constants $f_{K}=f_{\pi}$,
the magnetic moments of the decuplet baryons are simply given by\cite{lee} 
\begin{equation}
\mu_{B}=Q_{EM}(\frac{1}{16}{\cal M}+\frac{1}{4}({\cal N}-\frac{1}{2\sqrt{3}} 
{\cal N}^{\prime}))  \label{q}
\end{equation}
where $Q_{EM}$ is the EM charge. Here one notes that in the CBM in the
adjoint representation the prediction of the baryon magnetic moments with
the chiral symmetry is the same as that with the SU(3) flavor symmetry since
the mass-dependent term in ${\cal L}_{CSB}$ and ${\cal L}_{FSB}$ do not
yield any contribution to $J^{\mu}_{FSB}$ so that there is no terms with $%
{\cal P}$ and ${\cal Q}$ in (\ref{dec}).

Due to the degenerate d- and s-flavor charges in the SU(3) EM charge
operator $\hat{Q}_{EM}$, the CBM possesses the U-spin symmetry relations in
the baryon decuplet magnetic moments, similar to those in the octet baryons 
\cite{nrqm} 
\begin{eqnarray}
\mu_{\Delta^{-}}&=&\mu_{\Sigma^{*-}}=\mu_{\Xi^{*-}}=\mu_{\Omega^{-}} 
\nonumber \\
\mu_{\Delta^{0}}&=&\mu_{\Sigma^{*0}}=\mu_{\Xi^{*0}}  \nonumber \\
\mu_{\Delta^{+}}&=&\mu_{\Sigma^{*+}}  \label{uspin}
\end{eqnarray}
which are subset of the more strong symmetry relations (\ref{q}).

Since the SU(3) FSB quark masses do not affect the magnetic moments of the
baryon decuplet in the adjoint representation of the CBM, in the more
general SU(3) flavor symmetry broken case with $m_{u}=m_{d}\neq m_{s}$, $%
m_{\pi}\neq m_{K}$ and $f_{\pi}\neq f_{K}$, the decuplet baryon magnetic
moments with ${\cal P}$ and ${\cal Q}$ satisfy other sum rules\cite{decup} 
\begin{eqnarray}
\mu_{\Sigma^{*0}}&=&\frac{1}{2}(\mu_{\Sigma^{*+}}+\mu_{\Sigma^{*-}})
\label{sumrule1} \\
\mu_{\Delta^{-}}+\mu_{\Delta^{++}}&=&\mu_{\Delta^{0}}+\mu_{\Delta^{+}}
\label{sumrule2} \\
\sum_{B\in {\rm decuplet}}\mu_{B}&=&0.  \label{cg}
\end{eqnarray}
Also one can easily see that the CBM shares with the nonrelativistic quark
and chiral quark soliton models\cite{decup,kim} the following sum rules:%
\footnote{%
In fact, the baryon decuplet magnetic moments in the nonrelativistic quark
and chiral quark soliton models satisfy the model independent relations (\ref
{uspin}) and (\ref{cg}). Also the V-spin symmetry relations (\ref{vspins})
hold in the most general SU(3) flavor symmetry broken case of the CBM with
the higher representation mixing corrections (\ref{vspindel}).} 
\begin{eqnarray}
-4\mu_{\Delta^{++}}+6\mu_{\Delta^{+}}+3\mu_{\Sigma^{*+}}-6\mu_{\Sigma^{*0}}
+\mu_{\Omega^{-}}&=&0  \nonumber \\
-2\mu_{\Delta^{++}}+3\mu_{\Delta^{+}}+2\mu_{\Sigma^{*+}}-4\mu_{\Sigma^{*0}}
+\mu_{\Xi^{*-}}&=&0  \nonumber \\
-\mu_{\Delta^{++}}+2\mu_{\Delta^{+}}-2\mu_{\Sigma^{*0}} +\mu_{\Xi^{*0}}&=&0 
\nonumber \\
\mu_{\Delta^{++}}-2\mu_{\Delta^{+}}+\mu_{\Delta^{0}}&=&0  \label{sumrules}
\end{eqnarray}
and 
\begin{equation}
\mu_{\Delta^{0}}-\mu_{\Sigma^{*-}}=\mu_{\Sigma^{*+}}-\mu_{\Xi^{*0}} =\frac{1%
}{2}(\mu_{\Delta^{+}}-\mu_{\Xi^{*-}}) =\frac{1}{3}(\mu_{\Delta^{++}}-\mu_{%
\Omega^{-}}).  \label{vspins}
\end{equation}
Here one notes that the $\Sigma^{*}$ hyperons satisfy the identity $%
\mu_{\Sigma^{*}}(I_{3})=\mu_{\Sigma^{*0}}+I_{3}\Delta\mu_{\Sigma^{*}}$,
where $\Delta\mu_{\Sigma^{*}}=\frac{1}{16}{\cal M}+\frac{1}{4}({\cal N}-%
\frac{1}{2 \sqrt{3}}{\cal N}^{\prime})+\frac{3}{14}{\cal P}-\frac{5}{56}%
{\cal Q}$, such that $\mu_{\Sigma^{*+}}+\mu_{\Sigma^{*-}}$ is independent of 
$I_{3}$ as in (\ref{cg}). For the $\Delta$ baryons one can formulate the
relation $\mu_{\Delta}(I_{3})=\mu_{\Delta}^{0}+I_{3}\Delta\mu_{\Delta}$ with 
$\mu_{\Delta}^{0}=\frac{1}{32}{\cal M}+\frac{1}{8}({\cal N}-\frac{1}{2\sqrt{3%
}} {\cal N}^{\prime})+\frac{1}{7}{\cal P}+\frac{5}{112}{\cal Q}$ and $%
\Delta\mu_{\Delta}=\frac{1}{16}{\cal M}+\frac{1}{4}({\cal N}-\frac{1} {2%
\sqrt{3}}{\cal N}^{\prime})+\frac{4}{21}{\cal P}-\frac{11}{168}{\cal Q}$, so
that $\Delta$ baryons can be easily seen to fulfill the second sum rule in (%
\ref{cg}) and the last one in (\ref{sumrules}). Also the summation of the
magnetic moments over all the decuplet baryons vanish to yield the model
independent relation, namely the third sum rule in (\ref{cg}), since there
is no SU(3) singlet contribution to the magnetic moments as in the baryon
octet magnetic moments.

\section{Strange form factors in multi-quark structure}

\setcounter{equation}{0} \renewcommand{\theequation}{\arabic{section}.%
\arabic{equation}}

Until now we have considered the CBM in the adjoint representation where the
U-spin symmetry is broken only through the magnetic moment operators $\hat{%
\mu}^{i(a)}_{FSB}$ induced by the symmetry breaking derivative term. To take
into account the missing chiral symmetry breaking mass effect from ${\cal L}%
_{CSB}$ and ${\cal L}_{FSB}$, in this section we will treat perturbatively
the symmetry breaking mass terms via the higher dimensional IR channels
where, differently from the nonperturbative Yabu-Ando scheme\cite{yabu}, the
CBM can be handled in the quantum mechanical perturbation theory with the
higher IR mixing in the baryon wave function to yield the minimal
multi-quark structure\cite{multi}.

On quantizing the collective variable $A(t)$, one can obtain the Hamiltonian 
$H=H_{0}+H_{SB}$ where 
\begin{eqnarray}
H_{0}&=&M+\frac{1}{2}(\frac{1}{{\cal I}_{1}}-\frac{1}{{\cal I}_{2}})\hat{J}%
^{2} +\frac{1}{2{\cal I}_{2}}(\hat{C}^{2}_{2}-\frac{3}{4}\hat{Y}^{2}_{R}) 
\nonumber \\
H_{SB}&=&m(1-D_{88}^{8})  \label{h}
\end{eqnarray}
where ${\cal I}_{1}$ and ${\cal I}_{2}$ are the moments of inertia of the
CBM along the isospin and the strange directions respectively and $\hat{J}%
^{2}$ and $\hat{C}^{2}_{2}$ the Casimir operators in SU$_{R}$(2) and SU$_{L}$%
(3) groups and $m$ the inertia parameter denoting the symmetry breaking
strength. Minimizing the static mass $M$ the soliton profile function
satisfies the Skyrme equation of motion in the meson phase 
\begin{equation}
(x^{2}+2\sin^{2}\theta)\frac{\d^{2}\theta}{\d x^{2}}+2x\frac{\d\theta}{\d x}
+((\frac{\d\theta}{\d x})^{2}-1-\frac{\sin^{2}\theta}{x^{2}})\sin 2\theta
-\mu_{\pi}^{2}x^{2}\sin\theta=0
\end{equation}
with $\mu_{\pi}=m_{\pi}/ef_{\pi}$ and $x=ef_{\pi}r$, the dimensionless
quantities. Here one notes that the pion mass yields deviation from the
chiral limit profile so that the numerical results in the massive Skyrmion
theory can be worsened when one uses the experimental decay constant. Since $%
(m_{u}+m_{d})/m_{s}\approx m_{\pi}^{2}/m_{K}^{2}\approx 0.1$ we will neglect
the light quark and pion masses.

For the baryon decuplet with $Y_{R}=1$ and $J=\frac{3}{2}$ the possible
SU(3) representations of the minimal multi-quark Fock space are restricted
by the Clebsch-Gordan series ${\bf 10}\oplus{\bf 27}\oplus{\bf 35}$ so that
the baryon decuplet wavefunctions are described by $|B\rangle=
|B\rangle^{10}-C_{27}^{B}|B\rangle^{27}-C_{35}^{B}|B\rangle^{35}$ where the
representation mixing coefficients are given by\footnote{%
The algorithm for the Clebsch-Gordan decomposition of the tensor product of
the two IRs in the qqq$\bar{{\rm q}}$q is given by $({\bf 3}\otimes{\bf 3}%
\otimes{\bf 3})\otimes(\bar{{\bf 3}}\otimes{\bf 3}) =({\bf 1}\oplus{\bf 8}%
^{2}\oplus{\bf 10})\otimes({\bf 1}\oplus{\bf 8}) ={\bf 1}^{3}\oplus{\bf 8}%
^{8}\oplus{\bf 10}^{4}\oplus\bar{{\bf 10}}^{2} \oplus{\bf 27}^{3}\oplus{\bf %
35}$ where the superscript denotes the number of the different IR's with the
same dimension. Due to the baryon constraint $Y_{R}=1$ coming from the WZW
term, the spin-$\frac{1}{2}$ octet baryons are restricted to the IR's ${\bf 8%
}\oplus\bar{{\bf 10}}\oplus{\bf 27}$ and the spin-$\frac{3}{2}$ decuplet
baryons to ${\bf 10}\oplus{\bf 27}\oplus{\bf 35}$.} 
\begin{equation}
C_{\lambda}^{B}=\frac{_{\lambda}\langle B|H_{SB}|B\rangle_{10}}{E_{\lambda}
-E_{10}}  \label{coef}
\end{equation}
with the eigenvalues $E_{\lambda}$ and eigenfunctions $|B\rangle^{\lambda}=
\Phi_{B}^{\lambda}\otimes|{\rm intrinsic}\rangle$ of the eigen equation $%
H_{0}|B\rangle^{\lambda}=E_{\lambda}|B\rangle^{\lambda}$. Here $%
\Phi_{B}^{\lambda}$ is the collective wavefunction discussed above and the
intrinsic state degenerate to all the baryons is described by a Fock state
of the quark operator and classical meson configuration. Using the decuplet
wavefunctions with the higher representation mixing coefficients (\ref{coef}%
) the additional hyperfine structure of the magnetic moment spectrum in the
perturbative scheme is given by 
\begin{equation}
\delta\mu_{B}^{i}=-2\sum_{\lambda=27,35}\frac{_{10}\langle B|\hat{\mu}^{i}
|B\rangle_{\lambda}{}_{\lambda}\langle B|H_{SB}|B\rangle_{10}} {%
E_{\lambda}-E_{10}}  \label{delmub}
\end{equation}
up to the first order of $m$.

Then one can obtain the V-spin symmetry relations in the perturbative
corrections of the decuplet magnetic moments 
\begin{eqnarray}
\delta\mu_{\Delta^{++}}&=&\delta\mu_{\Omega^{-}}=m{\cal I}_{2} (\frac{5}{672}%
{\cal M}+\frac{1}{168}{\cal N}+\frac{5}{336\sqrt{3}}{\cal N}^{\prime}) 
\nonumber \\
\delta\mu_{\Delta^{+}}&=&\delta\mu_{\Xi^{*-}}=m{\cal I}_{2} (\frac{5}{112}%
{\cal M}-\frac{1}{21}{\cal N}-\frac{5}{336\sqrt{3}} {\cal N}^{\prime}) 
\nonumber \\
\delta\mu_{\Delta^{0}}&=&\delta\mu_{\Sigma^{*-}}=m{\cal I}_{2} (\frac{55}{672%
}{\cal M}-\frac{17}{168}{\cal N}-\frac{5}{112\sqrt{3}}{\cal N}^{\prime}) 
\nonumber \\
\delta\mu_{\Delta^{-}}&=&m{\cal I}_{2}(\frac{5}{42}{\cal M}-\frac{13}{84} 
{\cal N}-\frac{25}{336\sqrt{3}}{\cal N}^{\prime})  \nonumber \\
\delta\mu_{\Sigma^{*+}}&=&\delta\mu_{\Xi^{*0}}=m{\cal I}_{2} (-\frac{1}{224}%
{\cal M}+\frac{5}{168}{\cal N}+\frac{11}{336\sqrt{3}} {\cal N}^{\prime}) 
\nonumber \\
\delta\mu_{\Sigma^{*0}}&=&m{\cal I}_{2}(\frac{13}{336}{\cal M}-\frac{1}{28} 
{\cal N}-\frac{1}{168\sqrt{3}}{\cal N}^{\prime}).  \label{vspindel}
\end{eqnarray}
Here one notes that including the above implicit FSB effects one can have
the sum rules (\ref{sumrule1}) and (\ref{sumrule2}), and 
\begin{equation}
\mu_{\Delta^{++}}-\mu_{\Delta^-}=3(\mu_{\Delta^+}-\mu_{\Delta^0})
\end{equation}
which also holds in the chiral perturbation theory\cite{but}.

Now in the multiquark structure of the CBM with the chiral and SU(3) FSB,
the magnetic moments of baryon decuplet can be broken up into three parts 
\begin{equation}
\mu_B =\mu_{0,B}({\cal M}, {\cal N}, {\cal N}^{\prime}) +\delta\mu_{1,B}(%
{\cal P}, {\cal Q}) +\delta\mu_{2,B}(m{\cal I}_2)
\end{equation}
where the first term $\mu_{0,B}$ comes from the chiral symmetric
contribution, $\delta\mu_{1,B}$ is due to the explicit FSB and $%
\delta\mu_{2,B}$ is obtained from the implicit FSB in the representation
mixing as shown in (\ref{delmub}). In Table 1 the baryon decuplet magnetic
moments $\mu_B$ are predicted in the CBM with the bag radius $R\sim 0.6$ fm
corresponding to the magic angle $\theta (R)=\pi/2$, where the baryon number
is shared equally with both quark and meson phases\cite{rho} and the
numerical values of the inertia parameters are given by ${\cal M}=0.66$, $%
{\cal N}=6.00$, ${\cal N}^{\prime}=0.52$, ${\cal P}=1.11$, ${\cal Q}=1.27$
and $m{\cal I}_{2}=3.96$ as in the baryon octet case. In particular we
obtain $\mu_{\Delta^{++}}=1.29 \mu_{p}$ comparable to the experimental value 
$\mu_{\Delta^{++}}^{exp}=(1.62 \pm 0.18) \mu_{p}$\cite{boss}. For $%
\mu_{\Omega^{-}}$ the CBM seems to well predict the value $-1.75$ n.m.,
consistent with the experimental data $-1.94\pm 0.17 \pm 0.14$ n.m.\cite{die}
since $\mu_{\Omega^{-}}$ could be mainly achieved from the strange quark and
kaon whose masses are kept in our massless profile approximation $(m_u +m_d
)/m_s \approx m_{\pi}^2 /m_K^2 \approx 0$. Also one notes that the implicit
FSB effects with the V-spin symmetry improve the prediction of $%
\mu_{\Delta^{++}}$, but that of $\mu_{\Omega^{-}}$ seems worsened. In Table
2, the baryon decuplet magnetic moments obtained from other calculations
such as the nonrelativistic quark model, relativistic quark model\cite{sch},
lattice gauge theory\cite{lein}, chiral perturbation theory\cite{but} and
Skyrmion model\footnote{%
For the Skyrmion model corresponding to the CBM with vanishing bag radius,
we have used the numerical values of the inertia parameters, ${\cal M}=0.67$%
, ${\cal N}=5.03$, ${\cal N}^{\prime}=0.91$, ${\cal P}=0.76$, ${\cal Q}=0.99$
and $m{\cal I}_{2}=1.79$.}, are listed together with the experimental data.

Next in the SU(3) flavor symmetry broken case we decompose the EM currents
into three pieces $J^{\mu}=J^{\mu (u)}+J^{\mu (d)}+J^{\mu (s)}$ where the
q-flavor currents $J^{\mu (q)}=J^{\mu (q)}_{CS}+J^{\mu (q)}_{FSB}$ are given
by substituting the charge operator $\hat{Q}$ with the q-flavor charge
operator $\hat{Q}_{q}$ 
\begin{eqnarray}
J^{\mu (q)}_{CS}&=&\bar{\psi}\gamma^{\mu}\hat{Q}_{q}\psi\Theta_{B} +(-\frac{i%
}{2}f_{\pi}^{2}{\rm tr}(\hat{Q}_{q}l^{\mu})+\frac{i}{8e^{2}} {\rm tr}[\hat{Q}%
_{q},l^{\nu}][l^{\mu},l^{\nu}]+U\leftrightarrow U^{\dagger}) \bar{\Theta}_{B}
\nonumber \\
& &+\frac{N_{c}}{48\pi^{2}}\epsilon^{\mu\nu\alpha\beta}{\rm tr}(\hat{Q}_{q}
l_{\nu}l_{\alpha}l_{\beta}-U\leftrightarrow U^{\dagger}) \bar{\Theta}_{B} 
\nonumber \\
J^{\mu (q)}_{FSB}&=&-\frac{i}{12}f_{\pi}^{2}(\chi^{2}-1){\rm tr} ((1-\sqrt{3}%
\lambda_{8})(U\hat{Q}_{q}l^{\mu}+l^{\mu}\hat{Q}_{q}U^{\dagger})
+U\leftrightarrow U^{\dagger})\bar{\Theta}_{B}  \nonumber
\end{eqnarray}
to obtain the baryon decuplet magnetic moments in the s-flavor channel 
\begin{eqnarray}
\mu_{\Delta}^{(s)}&=&-\frac{7}{48}{\cal M}+\frac{1}{12}({\cal N} -\frac{1}{2%
\sqrt{3}}{\cal N}^{\prime})+\frac{2}{21}{\cal P} +\frac{5}{168}{\cal Q} 
\nonumber \\
& &+m{\cal I}_{2}(\frac{85}{2016}{\cal M}-\frac{25}{504}{\cal N}-\frac{5}{%
252 \sqrt{3}}{\cal N}^{\prime})  \nonumber \\
\mu_{\Sigma^{*}}^{(s)}&=&-\frac{1}{6}{\cal M}+\frac{1}{126}{\cal P} -\frac{1%
}{126}{\cal Q}  \nonumber \\
& &+m{\cal I}_{2}(\frac{13}{504}{\cal M}-\frac{1}{42}{\cal N} -\frac{1}{252 
\sqrt{3}}{\cal N}^{\prime})  \nonumber \\
\mu_{\Xi^{*}}^{(s)}&=&-\frac{3}{16}{\cal M}-\frac{1}{12}({\cal N} -\frac{1}{2%
\sqrt{3}}{\cal N}^{\prime})-\frac{2}{21}{\cal P} -\frac{5}{168}{\cal Q} 
\nonumber \\
& &+m{\cal I}_{2}(\frac{3}{224}{\cal M}-\frac{1}{168}{\cal N}+\frac{1}{168 
\sqrt{3}}{\cal N}^{\prime})  \nonumber \\
\mu_{\Omega}^{(s)}&=&-\frac{5}{24}{\cal M}-\frac{1}{6}({\cal N} -\frac{1}{2%
\sqrt{3}}{\cal N}^{\prime})-\frac{3}{14}{\cal P} -\frac{1}{28}{\cal Q} 
\nonumber \\
& &+m{\cal I}_{2}(\frac{5}{1008}{\cal M}+\frac{1}{252}{\cal N} +\frac{5}{504 
\sqrt{3}}{\cal N}^{\prime}).
\end{eqnarray}
Here one notes that in general all the baryon decuplet magnetic moments
fulfill the model independent relations in the u- and d-flavor components
and the I-spin symmetry of the isomultiplets with the same strangeness in
the s-flavor channel 
\begin{equation}
\mu_{B}^{(d)}=\frac{Q_{d}}{Q_{u}}\mu_{\bar{B}}^{(u)},~~~ \mu_{B}^{(s)}=\mu_{%
\bar{B}}^{(s)}
\end{equation}
with $\bar{B}$ being the isospin conjugate baryon in the isomultiplets of
the baryon.

Now the form factors of the decuplet baryons, with internal structure, are
defined by the matrix elements of the EM currents 
\begin{equation}
\langle p+q|J^{\mu}|p\rangle=\bar{u}(p+q)(\gamma^{\mu}F_{1B}(q^{2}) +\frac{i%
}{2m_{B}}\sigma^{\mu\nu}q^{\nu}F_{2B}(q^{2}))u(p)
\end{equation}
where $u(p)$ is the spinor of the baryons and $q$ is the momentum transfer.
Using the s-flavor charge operator in the EM currents as before, in the
limit of zero momentum transfer, one can obtain the strange form factors of
baryon decuplet 
\begin{eqnarray}
F_{1B}^{(s)}(0)&=&S  \nonumber \\
F_{2B}^{(s)}(0)&=&-3\mu_{B}^{(s)}-S
\end{eqnarray}
in terms of the strangeness quantum number of the baryon $S(=1-Y)$ ($Y$:
hypercharge) and the strange components of the baryon decuplet magnetic
moments $\mu_{B}^{(s)}$.

As shown in Table 3, we have obtained the theoretical predictions of the CBM
and the Skyrmion model for the strange form factors of baryon octet and
decuplet. Here one notes that the implicit FSB contributions to the strange
form factors in the CBM are quite significant in the $\Delta$ and $\Sigma^{*}
$ decuplet baryons. Also the CBM prediction $F_{2p}^{(s)}=0.30$ n.m. for the
proton strange form factor is comparable to the SAMPLE Collaboration
experimental result $F_{2p}^{(s),exp}= 0.23 \pm 0.37 \pm 0.15 \pm 0.19$ n.m.
\cite{sample}.

\section{Conclusions}

\setcounter{equation}{0} \renewcommand{\theequation}{\arabic{section}.%
\arabic{equation}}

In this paper, we have considered the strange form factors of the baryon
decuplet by including the implicit FSB effects via the higher dimensional IR
mixing scheme. In this approach the CBM can be treated to yield the minimal
multiquark structure in the quantum mechanical perturbation theory.

Using the chiral bag with the bag radius $R\sim 0.6$ fm corresponding to the
magic angle $\theta (R)=\pi/2$ we have obtained the magnetic moments $%
\mu_{\Delta^{++}}=3.59$ n.m. and $\mu_{\Omega^-}=-1.75$ n.m., comparable to
the experimental values $\mu_{\Delta^{++}}^{exp}=4.52\pm 0.50$ n.m. and $%
\mu_{\Omega^-}^{exp}=-1.94\pm 0.17 \pm 0.14$ n.m., respectively. Here one
notes that the proton strange form factor was predicted\cite{hpm} in the CBM
with the value $F_{2p}^{(s)}=0.30$ n.m. which is also comparable to the
SAMPLE Collaboration experimental data $F_{2p}^{(s),exp}=0.23 \pm 0.37 \pm
0.15 \pm 0.19$ n.m.\cite{sample}.

From the SU(3) group structure of the chiral models, such as Skyrmion, MIT
and chiral bag models, we have derived the sum rules among the magnetic
moments of the baryon decuplet. Here one notes that these model independent
sum rules also hold in the nonrelativistic quark and chiral quark soliton
models and the chiral perturbation theory.

Substituting the charge operator in the EM currents with the s-flavor charge
operator, we have obtained, in the zero momentum transfer limit, the strange
form factors of the baryon decuplet in terms of the strangeness quantum
number of the baryon and the strange components of the baryon decuplet
magnetic moments. Here one notes that the strange form factors are
degenerate in the isomultiplets with the same strangeness to respect the
I-spin symmetry.

On the other hand, in the multiquark structure of the CBM with the chiral
and SU(3) FSB, the strange form factors of baryon decuplet can be broken up
into three parts: the chiral symmetric contribution, the explicit FSB one
and the implicit FSB one with the representation mixing. The theoretical
predictions for the strange form factors show that the implicit FSB effects
are dominant in the $\Delta$ and $\Sigma^*$ decuplet baryons and in any case
the higher dimensional IR mixing in the baryon wave function in the
multiquark Fock space cannot be omitted in the strange form factors of the
decuplet baryons.

\vskip 1cm We would like to thank B.Y. Park, M. Rho and G.E. Brown for
helpful discussions and constant concerns. This work is supported in part by
the Korea Science and Engineering Foundation through the CTP and by the
Korea Ministry of Education under Grant No. BSRI-98-2418.

\newpage 
%------------------------------------------------------------------------
\begin{table}[t]
\caption{The baryon decuplet magnetic moments $\protect\mu_{B}=\protect\mu%
_{0,B} +\protect\delta\protect\mu_{1,B}+\protect\delta\protect\mu_{2,B}$
calculated in the chiral bag with bag radius $R\sim 0.6$ fm corresponding to
the magic angle $\protect\theta (R)=\protect\pi/2$.}
\begin{center}
\begin{tabular}{lrrrr}
\hline
$B$ & $\mu_{0,B}$ & $\delta\mu_{1,B}$ & $\delta\mu_{2,B}$ & $\mu_{B}$ \\ 
\hline
$\Delta^{++}$ & $3.00$ & 0.41 & $0.18$ & $3.59$ \\ 
$\Delta^{+}$ & $1.50$ & 0.28 & $-1.03$ & $0.75$ \\ 
$\Delta^{0}$ & $0.00$ & $0.15$ & $-2.24$ & $-2.09$ \\ 
$\Delta^{-}$ & $-1.50$ & $0.02$ & $-3.45$ & $-4.93$ \\ 
$\Sigma^{*+}$ & $1.50$ & $0.12$ & $0.73$ & $2.35$ \\ 
$\Sigma^{*0}$ & 0.00 & 0.00 & $-0.76$ & $-0.76$ \\ 
$\Sigma^{*-}$ & $-1.50$ & $-0.13$ & $-2.24$ & $-3.87$ \\ 
$\Xi^{*0}$ & $0.00$ & $-0.15$ & $0.73$ & $0.58$ \\ 
$\Xi^{*-}$ & $-1.50$ & $-0.28$ & $-1.03$ & $-2.81$ \\ 
$\Omega^{-}$ & $-1.50$ & $-0.43$ & $0.18$ & $-1.75$ \\ \hline
\end{tabular}
\end{center}
\end{table}
%------------------------------------------------------------------------
\begin{table}[t]
\caption{The baryon decuplet magnetic moments $\protect\mu_{B}^{CBM}$
and $\protect\mu_{B}^{SM}$ calculated in the chiral bag model (CBM) and 
Skyrmion model (SM), compared with the nonrelativistic quark model (NRQM), 
relativistic quark model (RQM), lattice gauge theory (LGT), chiral perturbation 
theory (CPT) and the experimental data.${}^{\dagger}$  The quantity used 
as input is indicated by $*$.}
\begin{center}
\begin{tabular}{lrrrrrr}
\hline
$B$ & $\mu_{B}^{NRQM}$ & $\mu_{B}^{RQM}$ & $\mu_{B}^{LGT}$ & $\mu_{B}^{CPT}$
& $\mu_{B}^{SM}$ & $\mu_{B}^{CBM}$ \\ \hline
$\Delta^{++}$ & $5.58$ & 4.76 & $4.91$ & 4.00 & 2.82 & $3.59$ \\ 
$\Delta^{+}$ & $2.79$ & 2.38 & $2.46$ & 2.10 & 1.04 & $0.75$ \\ 
$\Delta^{0}$ & $0.00$ & $0.00$ & $0.00$ & $-0.17$ & $-0.74$ & $-2.09$ \\ 
$\Delta^{-}$ & $-2.79$ & $-2.38$ & $-2.46$ & $-2.25$ & $-2.52$ & $-4.93$ \\ 
$\Sigma^{*+}$ & $3.11$ & $1.82$ & $2.55$ & 2.00 & 1.60 & $2.35$ \\ 
$\Sigma^{*0}$ & 0.32 & $-0.27$ & $0.27$ & $-0.07$ & $-0.28$ & $-0.76$ \\ 
$\Sigma^{*-}$ & $-2.47$ & $-2.36$ & $-2.02$ & $-2.20$ & $-2.17$ & $-3.87$ \\ 
$\Xi^{*0}$ & $0.64$ & $-0.60$ & $0.46$ & 0.10 & 0.18 & $0.58$ \\ 
$\Xi^{*-}$ & $-2.15$ & $-2.41$ & $-1.68$ & $-2.00$ & $-1.81$ & $-2.81$ \\ 
$\Omega^{-}$ & $-1.83$ & $-2.35$ & $-1.40$ & $-1.94^{*}$ & $-1.46$ & $-1.75$
\\ \hline
\end{tabular}
\end{center}
\par
{${}^{\dagger}$ For the experimental data $\mu_{\Delta^{++}}^{exp}=4.52\pm 
0.50$ and $\mu_{\Omega^{-}}^{exp}=-1.94\pm 0.17 \pm 0.14$ we have referred to 
the ref. \cite{boss} and ref.\cite{die}, respectively.}\par
\end{table}

\newpage 
%------------------------------------------------------------------------
\begin{table}[t]
\caption{The strange form factors of the baryon octet and decuplet $%
F_{2B}^{(s)}=F_{2B}^{(s),0}+\protect\delta F_{2B}^{(s),1}+\protect\delta %
F_{2B}^{(s),2}$ calculated in the chiral bag, compared with the Skyrmion
model predictions $F_{2B}^{(s)SM}$.}
\begin{center}
\begin{tabular}{lrrrrr}
\hline
$B$ & $F_{2B}^{(s),0}$ & $\delta F_{2B}^{(s),1}$ & $\delta F_{2B}^{(s),2}$ & 
$F_{2B}^{(s)}$ & $F_{2B}^{(s)SM}$ \\ \hline
$N$ & $-0.19$ & $-0.12$ & $0.61$ & $0.30$ & $-0.02$ \\ 
$\Lambda$ & $0.55$ & $0.35$ & $-0.41$ & $0.49$ & 0.51 \\ 
$\Sigma$ & $-1.89$ & $-0.34$ & $0.69$ & $-1.54$ & $-1.74$ \\ 
$\Xi$ & $0.07$ & $0.45$ & $-0.27$ & $0.25$ & 0.09 \\ 
$\Delta$ & $-1.17$ & $-0.43$ & $3.27$ & $1.67$ & 0.04 \\ 
$\Sigma^{*}$ & $-0.67$ & $0.00$ & $1.51$ & $0.84$ & $-0.10$ \\ 
$\Xi^{*}$ & $-0.17$ & $0.43$ & $0.30$ & $0.56$ & $-0.03$ \\ 
$\Omega$ & $0.34$ & $0.85$ & $-0.36$ & $0.83$ & 0.24 \\ \hline
\end{tabular}
\end{center}
\end{table}
%------------------------------------------------------------------------
\end{document}